# About_the_electrodynamic_acceleration_of_cylinder-shaped_particles

*S. N. Dolya*

*Joint Institute for Nuclear Research, ul. Joliot – Curie 6, Dubna, 141980 Russia*

**Abstract**

A possibility of electrodynamic acceleration of particles from the initial zero velocity to the final velocity $V_{fin}$ = 10 km/s over the acceleration length $L_{acc}$ = 5 m is considered. After the electrostatic preacceleration to $V_{in}$ = 1 km/s particles are accelerated at the trailing edge of the voltage pulse $U_{pulse}$ = 5.75 MV, which runs along the spiral turns. Accelerated particles have the mass m = 6 * $10^{-6}$ g, diameter $d_{sh}$ = 6 μ, and length $l_{sh}$ = 1 cm. Because of a pointed cone at the head the particles can move in the air almost without loss of velocity, penetrating into aluminum and water as deep as $l_{Al}$ = 10 cm and $l_{water}$ = 1 m respectively.

**Introduction**

It is known [1] that magnetic dipoles can be accelerated by orienting them in space so that the magnetization axis coincides with the acceleration axis and applying the current pulse field.

Being a difference interaction, the dipole interaction is inherently much less effective than the monopole charge interaction. For example, if one of the poles of the dipole is repelled by the accelerating pulse, the other is at the time is attracted to the current pulse and the resulting accelerating force is the difference of these two forces. Therefore, a uniform magnetic field does not act on the dipoles; the dipoles can be accelerated by the magnetic field gradient.

In addition, iron, which is the most suitable material for acceleration, has a low specific magnetic moment (magnetic moment per nucleon):
m = 2 * $10^{-10}$ eV / (Gs * nucleon). Accordingly, the energy gain rate of the iron magnetic dipole: m * ∂ B / ∂ z in a gradient magnetic field with a reasonable gradient ∂ B / ∂ z = 500 Gs / cm is found to be
$\Delta W_{mag}$ = m * ∂ B / ∂ z = 2 * $10^{-10}$ * 500 = $10^{-7}$ eV / cm.

Energy gain by a particle having a specific charge (the charge per nucleon in units of electron charge) can amount to e (Z / A) = $10^{-8}$, and in the electric field E = 10 kV / cm the energy gain rate is found to be
$\Delta W_{el}$ = $10^{-8}$ * $10^{4}$ = $10^{-4}$ eV / cm, which is three orders of magnitude higher than the energy gain rate in the dipole interaction.

The spherical shape of the particles used in some systems is not optimal for many reasons, especially acceleration-related ones. As the radius of particle increases, the acceleration efficiency dramatically drops. To understand better the fundamental shortcomings of acceleration of spherical particles, let us compare the main parameters of accelerated iron balls with different diameter.

Table 1. The main parameters of the accelerated objects

| D, μ | A | Z | Z/A | eΦ, MeV | M, g | $\beta_i$ |
|---|---|---|---|---|---|---|
| 2 | $2*10^{13}$ | $6*10^7$ | $3*10^{-6}$ | 0.1 | $3.2*10^{-11}$ | $4*10^{-5}$ |
| 20 | $2*10^{16}$ | $6*10^9$ | $3*10^{-7}$ | 1 | $3.2*10^{-8}$ | $1.2*10^{-5}$ |
| 200 | $2*10^{19}$ | $6*10^{11}$ | $3*10^{-8}$ | 10 | $3.2*10^{-5}$ | $4*10^{-6}$ |
| $2*10^3$ | $2*10^{22}$ | $6*10^{13}$ | $3*10^{-9}$ | $10^2$ | $3.2*10^{-2}$ | $1.2*10^{-6}$ |

In all cases the electric field on the surface of the object is $E_{surf} = 10^9$ V / cm. The first column is the diameter of the ball in μm, the second column is the atomic weight of the ball in units of the atomic mass of the nucleon, the third column is the limiting charge Z on the object in units of the electron charge, the fourth column is the ratio of the charge on the ball to its weight, the fifth column is the potential Φ of the object (the energy that an electron must have to overcome the repulsion of the electrons previously placed on the object), the sixth column is the mass of the ball in grams, and the seventh column is the initial velocity of the balls acquired by them after the acceleration in the electrostatic field with the voltage $U_{inj} = 250$ kV, expressed in units of the speed of light $\beta_{in}$ = V / c, where c = 3 * $10^5$ km / s is the velocity of light in vacuum.

A comparison of the data presented in Table 1 shows that as the diameter of the balls increases, their weight and atomic mass (column 6) increase as the cube of the radius, and the charge to be placed on the ball to get the field strength $E_{surf} = 10^9$ V / cm increases as the square of the radius. For a ball with the diameter D = 2 mm the charge (in units of electron charge) Q = I * τ = 2 A * 5 μs = $6 * 10^{13}$ is already close to the limiting charge per pulse accelerated in linear accelerators. The ratio of the charge placed on the ball to the mass of the ball (column 4) linearly decreases with increasing diameter, which means that the acceleration efficiency linearly with increasing diameter, i.e., balls of a larger diameter will gain less velocity over the same accelerator length and at the same intensity.

For cylinder-shaped particle the parameter Z / A does not depend on the length of the cylinder at all because the charge placed on the particle and its mass linearly increase with increasing length of the cylinder. Platinum coating and oxygen passivation of the particle greatly increases its surface barrier preventing electrons from leaking from with particles.

The cylindrical shape of particles with a pointed cone at the head allows much better ballistic performance than that of the prototype, namely, the pointed cone ensures a very low drag coefficient, which is important for moving through the atmosphere without losing velocity. The use of tungsten instead of iron significantly improves the aerodynamic "quality" of particles.

Finally, elongated cylindrical particles will be able to penetrate into a material many time deeper than spherical particles, which is of interest for some practical applications.

**Accelerator**

*1. Introduction of 3.3% of iron in tungsten*

Let us add 3.3% of iron to a tungsten particle and magnetize this macroparticle so that the axis of magnetization coincides with the longitudinal axis of the cylinder and the axis of acceleration. Since the atomic weight of tungsten is about 3.3 times greater than that of iron and tungsten does not has intrinsic magnetic moment, the resulting specific magnetic moment will be approximately 100 times lower than that of iron:
$m_1 = 2 * 10^{-12}$ eV / (Gs * nucleon).

*2. Platinum coating and oxygen passivation of the tungsten wire*

To create a surface barrier for electrons "accommodated" on the particle, the work function of the electrons must be increased as much as possible. Best-known is the work function of platinum passivated with oxygen [2].
The charge located on the macroparticle will leak from it through field emission according to the formula [2],

$$j = e^2E^2 / (8\pi h\varphi) * \exp \{[- (8\pi / 3) (2m)^{1/2}/h] * [(e\varphi)^{3/2} / (eE) * \theta (y)]\}, \quad (1)$$

where $\theta (y)$ is the Nordheim function in which the argument is a relative reduction in the work function of an external electric field in the sense of

Schottky.

*3. Spatial orientation and pre-acceleration of particles*

We estimate the velocity of particles accelerated by a coil in which the current is turned off when the macroparticle reaches the middle of the coil. The field strength on the axis of the coil is described by the formula

$$B_z = [2\pi I_\varphi / c] * R_0^2 / [z^2 + R_0^2]^{3/2}, \qquad (2)$$

where $R_0$ is the radius of the current coil turn, $I_\varphi$ is the current in the turn, and z is the distance from the center of the turn.

Let the particle be pre-magnetized and have a specific magnetic moment two orders of magnitude lower than that of iron, namely,
$m_1 = 2 * 10^{-12}$ eV / (Gs * nucleon). Using formula (2), we find the magnetic field gradient $\partial B_z / \partial z = B_z * 3z / [z^2 + R_0^2]$ and the particle energy increment resulting from the interaction of the particle magnetic moment with the magnetic field gradient of the current turn

$$\Delta W_{in1} = m_1 * (\partial B_z / \partial z) * l_{acc}, \qquad (3)$$

which for $l_{acc} = 10$ cm and the magnetic field gradient
$(\partial B_z / \partial z) = 500$ Gs / cm can be estimated at $\Delta W_{in1} = 10^{-8}$ eV / nucleon, so that the velocity of the particles after pre-acceleration in this turn (coil gun) will be $V_{in1} = 1.5$ m / s and the segment will be directed along the axis of acceleration.

*4. Electron beam irradiation of particles: Electron energy*

Thus, we have slightly accelerated the macroparticle due to its weak magnetization and, which is important, oriented it along the axis of acceleration. Further acceleration of the particle will be due to its interaction with the electric field; therefore, it must be an electrically charged macroparticle. The "charge" can be produced by irradiating the macroparticle with an electron beam.

The surface strength of the field is takes to be $E_{surf} = 40$ MV / cm (later we will analyze in detail the choice of this very important parameter). Then, for the wire diameter $d = 6$ µ we find that the minimum energy of the electrons capable of overcoming the Coulomb repulsion of the electrons previously placed on the macroparticle should be $W_e > eE_{surf} * d_{sh} / 2 = 12$ keV.

*5. Electron beam irradiation of particles: Mean free path*

The path of electrons with an energy of 12 keV in aluminum is approximately 2 mg/cm² [2]. Given the density of aluminum $\rho_{Al} = 2.7$ g/cm³, we find that the extrapolated path of electrons in aluminum is $l_{Al} \approx 7$ μ. Since the tungsten density is $\rho_{tung} = 19.35$ g/cm³, more than 7 times higher than the density of aluminum, the mean free path of electrons with an energy of 12 keV in tungsten is approximately 1 μ. Thus, electrons with the energy $W_e = 12$ keV are not able to cross the particle diameter 6 μ, will lose their energy by ionization of matter, and will stop deep in the particle.

*6. Electron beam irradiation of particles: Irradiating beam density*

We estimate the time of flight of the macroparticle over a distance equal to its own length, that is, $l_{sh} = 1$ cm. At the particle velocity $V_{sh} = 1.5$ m / s this time will be $t_{sh} = l_{sh} / V_{sh} = 7$ ms. Let the duration of the electron beam irradiation be $t_{irr} = 300$ μs, Below we will show that during this time the macroparticle fully charged electrons will lose less than one percent of electrons through field emission. Let the total number of electrons placed on the particle be $N_e = 4 * 10^{10}$.

Let the transverse dimensions of the irradiating beam ribbon be $S_{irr} = 5μ * 1cm$, where the beam cross section 5μ is chosen such that the length of the chord drawn through the cross section of the particle is greater than 1 μ, the depth of the extrapolated path of the electrons with the energy $W_e = 12$ keV in tungsten. Considering that the reflectance of slow electrons from tungsten is about $k_{irr} = 0.35$ [2], we find from the relation

$$j_{irr} * 6 * 10^{18} * S_{irr} * t_{irr} * k_{irr} = N_e \qquad (4)$$

that the desired current density of the electron beam irradiation is approximately $j_{irr} = 0.1$ A/cm², which is much lower than, for example, the electron emission current density obtained from oxide cathodes.

*7. Electron beam irradiation of particles: Introduced pulse*

We estimate the transverse particle velocity arising from the pulse introduced by electrons. The velocity of the electrons irradiating the particle is $\beta_e = V_e / c = 5 * 10^{-3}$, where $c = 3 * 10^{10}$ is the speed of light in vacuum. The transverse velocity of the particles resulting from electron irradiation will be

$$V \perp = (N_e / A_a N_a) * m_e V_e / M_n, \qquad (5)$$

where $N_e$ is the number of uncompensated (planted on macroparticle) electrons, $N_a$ is the number of atoms in the particle, $A_a$ is the atomic weight of tungsten, $m_e$ is the electron mass, and $M_n$ is the mass of the nucleon. If there are few excess electrons in the macroparticle $(N_e / A_a N_a) = (Z / A) = 1.23 * 10^{-8}$ and the electron-to-nucleon mass ratio is 1/2000, the transverse velocity is low. If it is a problem, particles will have to be irradiated with the two colliding electron beams.

*8. Electron beam irradiation of particles: Field electron emission*

To plant several charges on a macroparticle is not a problem, but when there are many electrons on the macroparticle, they will begin to leak from it through field emission. Let the field strength for field emission be $E = 4 * 10^7$ V / cm. After this field strength is attained, all the newly planted electrons will leak from the particles due to the Coulomb repulsion.

Once there is quite a lot of electrons planted on the particle, it becomes necessary to overcome their repulsion for planting more electrons. This means that the energy of electrons that we want to plant on the particle should be high enough to allow them to overcome this Coulomb barrier, reach the particle, and stay on it.

For spherical particles the Coulomb barrier linearly increases with increasing radius, and for particles with the diameter d = 2 mm it can be as high as tens of MeV, which will require a special accelerator for the electrons to overcome it.

Field emission will be an obstacle for planting a high electric charge on the particle. Part of the charge will continuously leak from particles due to the tunnel effect. The dependence of the field-emission current density on the electric field strength for surfaces with different work functions is given in [2].

We take the following parameters: density $\rho_{tung} = 19$ g/cm$^3$, particle diameter $d_{sh} = 6$ μ, length $l_{sh} = 1$ cm, particle cross-section $S_{tr} = \pi d^2 / 4 = 2.8 * 10^{-7}$ cm$^2$, volume $V_{sh} = 2.8 * 10^{-7}$ cm$^3$, the mass of the tungsten wire segment (particle) $m_{sh} = 6 * 10^{-6}$ g, lateral particle surface $S_{surf} = \pi d_{sh} * l_{sh} = 10^{-3}$ cm$^2$, and surface tension of the field on the macroparticle $E_{surf} = 4 * 10^7$ V / cm.

Let us find the charge on the particle that is enough to create the electric field $E_{surf} = 4 * 10^7$ V / cm from the expression relating the strength of the field on a cylinder to the linear charge density on it

$$E_{surf} = 2\kappa / r = 4 * 10^7 = 2eN_e * 300 / r, \qquad (6)$$

Thus, $N_e = 4 \times 10^{10}$.

*9. Acceleration of particles by the electrostatic field*

Let us now calculate the resulting ratio Z / A. The proportion for tungsten is

$$6 * 10^{23} \text{ nucleons -} \quad 184 \text{ g}$$

$$N_{tung\ nucleons} - \quad 19.35 \text{ g.} \qquad (7)$$

We find from it that in one cubic centimeter of tungsten there are $N_{tung\ nucleons} = 1.16 * 10^{25}$ nucleons. In the wire segment with the diameter $d_{sh} = 6$ μ and length $l_{sh} = 1$ cm there are $3.25 * 10^{18}$ nucleons, and the ratio Z / A for this segment is

$$Z / A = 4 * 10^{10} / 3.25 * 10^{18} = 1.23 * 10^{-8}. \qquad (8)$$

Let the accelerating electrostatic field be $E_{acc} = 10$ kV / cm. The equation of motion of particles with a specific (per nucleon) charge $Z / A = 1.23 * 10^{-8}$ can be written as

$$dV / dt = (Z / A)\, eE_{acc} / M_n, \qquad (9)$$

where $M_n$ is the mass of the nucleon. Assuming that the initial particle velocity is zero, we obtain an expression for the time dependence of the dimensionless velocity $\beta = V / c$, where $c = 3 * 10^5$ km / s is the speed of light in vacuum, in the form

$$\beta = \{(Z / A) \, eE_{acc}/M_n c^2\} * ct. \quad (10)$$

Let the final acceleration velocity be $V_{fin} = 1$ km / s, which corresponds to the dimensionless velocity $\beta_{fin} = 3.3 * 10^{-6}$. From equation (10) we can find that $ct = 2.7 * 10^7$. Therefore, the time of acceleration from the initial zero velocity to the final velocity $V_{fin} = 1$ km / s is $t_{acc} = 9 * 10^{-4}$ s.

The acceleration length $L_{acc1}$ can be found from the formula of uniformly accelerated motion, which in our case can be written as

$$L_{acc1} = \{(Z / A) \, eE_{acc}/M_n c^2\} * (ct_{acc})^2/2. \quad (11)$$

Substituting numbers into (11), we find that the acceleration length is $L_{acc1} = 45$ cm; accordingly, the electrostatic voltage required for the acceleration of particles is $U_{acc}$

$$U_{acc} = E_{acc} * L_{acc1} = 450 \text{ kV}. \quad (12)$$

*10. Leakage of electrons*

Let us find the number of electrons that left the particle during the acceleration. Given the field strength $E = 40$ MV / cm and the work function $e\varphi = 6.5$ eV, we find from the graph [2] that the leakage current density is $j = 10^{-4}$ A/cm$^2$.

The charge leakage $\Delta Q$ will be:

$$\Delta Q = j * S_{surf} * t_{acc}, \quad (13)$$

where $j = 10^{-4}$ A/cm$^2$ is the leakage current, $S_{surf} = 1.88 * 10^{-3}$ cm$^2$ is the lateral surface area of the particle, and $t_{acc} = 9 * 10^{-4}$ s is the acceleration time. Substituting numbers into (13), we obtain $\Delta Q = 10^9$ electrons, which is about 2.5% of the number of electrons planted on the particle.

Gold-coated tungsten wire 5 to 500 µm in diameter has been produced for decades by the Swedish company LUMA [3].

*11. Electrodynamic acceleration*

A charged tungsten wire segment can be accelerated from the initial velocity $V_{in} = 1$ km / s to the final velocity $V_{fin} = 10$ km / s in a spiral waveguide with a voltage pulse running in its turns.

The final velocity $\beta_{fin} = 3.3 * 10^{-5}$ corresponds to the energy $W_{fin} = Mc^2 \beta^2_{fin} / 2 = 0.54$ eV / nucleon. At the accelerating rate $E_0 \sin\varphi_s = 100$ kV / cm this energy can be obtained over the length $L_{acc} = W_{fin} / [(Z / A) E_0 \sin\varphi_s] = 4.5$ m.

Taking the synchronous phase $\varphi_s = 60^0$, $\sin\varphi_s = 0.87$, we find that the amplitude of the electric field strength in a spiral waveguide must be $E_0 = 115$ kV / cm. It is not necessary to use a sine wave for accelerating a wire segment. Instead, we can use a single sine pulse equal to the half-cycle wave in time and half-length of the wave propagating in the spiral waveguide [4].

The relationship between flux and the electric field strength for a spiral with the space between the spiral and the screen filled with a dielectric is given by the formula [4]

$$P = (c / 8) r_0^2 E_0^2 \beta\varepsilon \{ \}, \qquad (14)$$

where $c = 3 \times 10^{10}$ cm / s is the speed of light in vacuum, $r_0 = 50$ cm is the initial radius of the spiral, $E_0 = 115$ kV / cm is the field strength on the axis of the spiral, $\beta$ is the phase velocity, and $\varepsilon = 1280$ is the dielectric constant of the filling between the spiral and the screen; the value of the brackets { } for the argument $x = 1$ is approximately 4.

The argument of the modified Bessel functions of the first and second kind, $x = 1 = 2\pi r_0 /\beta\lambda_0$, where $\beta$ is the phase velocity and $\lambda_0 = c/f_0$ is the vacuum wavelength, is optimal [4]. In our case, at the beginning of acceleration $r_0 = 50$ cm, $\beta = 3.3 * 10^{-6}$, $f_0 = 160$Hz, and $T_0 / 2 = 3.12$ ms. Substituting numbers into (14), we obtain

$$P (W) = (c / 8) * 2.5 * 10^3 * (1.15)^2 * 10^{10} * 3.3 * 10^{-6} * 1.28 * 10^3 * 4 / (9 * 10^4 * 10^7) = 2.32 \text{ GW}. \qquad (15)$$

The length of the slowed-down wave at the beginning of the acceleration is $\lambda_{slow} = \beta\lambda_0 = 3.14$ m, and the spatial length of the accelerating pulse is equal to half the wavelength $\Delta l_{pulse} = 1.57$ m. The base duration of the pulse is $T_0 / 2 = 3.12$ ms. The amplitude of the pulse can be evaluated from the relation

$$U_{pulse} = E_0 * \lambda_{slow}/2\pi = 5.75 \text{ MV}. \qquad (16)$$

From the relation $x = 1 = 2\pi r_0/\beta\lambda_0$ it follwos that for each phase velocity of the wave propagating in the spiral waveguide there is its own optimum spiral radius, optimum wavelength, frequency, and pulse duration. As it is quite difficult to cover an order-of-magnitude change in of the phase velocity from $V_{in} = 1$ km / s to $V_{fin} = 10$ km / s with one pulse duration value, it is reasonable to divide the spiral waveguide into sectionsin which a pulse of optimum duration propagates [4]. Since this pulse defocuses the accelerated particle, there should be focusing elements placed between the sections. In our case, electrostatic quadrupole lenses are most suitable [4].

*12. Release of particles into the atmosphere*

Particles must be irradiated with an electron beam and then accelerated in a rather high vacuum $P_1 \approx 10^{-9}$ mm Hg, while they are expected to be used under normal atmospheric conditions $P_2 \approx 10^3$ mm Hg, which gives the pressure difference of about 12 orders of magnitude. This pressure gradient can be produced using several buffer cavities, for example, cylindrical chambers separated by end walls with pulse diaphragms built into them. Each cavity must be individually evacuated.

Let us calculate the number of the air particles that penetrate the first cavity, nearest to the atmosphere. Let the radius of the diaphragm be $r_{0d} = 1$ cm and the time for which it remains open be $t_1 = 10^{-4}$ s. Then the linear velocity of the iris diaphragm petals will be $V_{11} = r_{0d} / t_1 = 10^4$ cm / s, which should not be a problem for the operation of the mechanism. The average velocity of the thermal motion of air molecules V can be assumed to be equal to the speed of sound in the air $V = 3 * 10^4$ cm / s; of all possible spatial orientations of the velocity only 1/6 (1/6 is one face of the cube) is directed towards the diaphragm. The number of molecules in a cubic centimeter of the air under normal conditions, the Loschmidt number, of is $\rho_{0L} = 2.7 * 10^{19}$ molecules/cm$^3$. Then the number of molecules coming from the atmosphere to the first buffer cavity while the diaphragm is open is

$$N_0 = (1/6) \rho_{0L} * \pi r_{0d}^2 * V_{t1}. \qquad (17)$$

Substituting numerical values into (14), we find that the total number of particles coming from the atmosphere to the first buffer cavity is $N_0 = 4 * 10^{19}$. Let the volume of the first cavity be $V_{01} = 10 \, l = 10^4 \, cm^3$. Then the density of molecules in it after the diaphragm operated will be $n_0 = 4 * 10^{15}$ molecules/$cm^3$.

Particle density (and pressure $p = nkT$) in the first buffer cavity is about 4 orders of magnitude lower than the density of particles in the atmosphere under normal conditions. Thus, at least three cavities are needed to produce the appropriate pressure gradient.

We now consider the dynamics of the particle density in the cavity for the time between the operations of the device. Let the cavity be pumped through the hole with an area $S_1 = 10^3 \, cm^2$. We assume that all molecules within this area are removed from the volume and that the time between the shots is $t_2 = 10^{-2}$ s, i.e., the operation frequency of the device is $F = 100$ Hz. The equation describing the particle density reduction during the pumping can be written as

$$dn = -n * (1/6) * S_1 V_t / V_{01}. \qquad (18)$$

The solution of this equation can be written as:

$$n = n_0 * \exp[-(1/6) * S_1 V_t / V_{01}] \qquad (19)$$

For the pumping time $t = t_2$ the exponent is approximately 5, and thus the pumping reduces the density of molecules in the first buffer cavity by a factor of more than 100. Then, $n = n_0 * 7 * 10^{-3}$, and the density of particles in the first buffer volume before the next shot will be $n_1 = 4 * 10^{15} * 7 * 10^{-3} = 3 \times 10^{13}$ molecules/$cm^3$, i.e., 6 orders of magnitude lower than the Loschmidt number $\rho_{0L} = 2.7 * 10^{19}$ molecules/$cm^3$.

It is evident that before the next shot the cavity can be considered empty.

Figure 1 shows the diagram of the device.

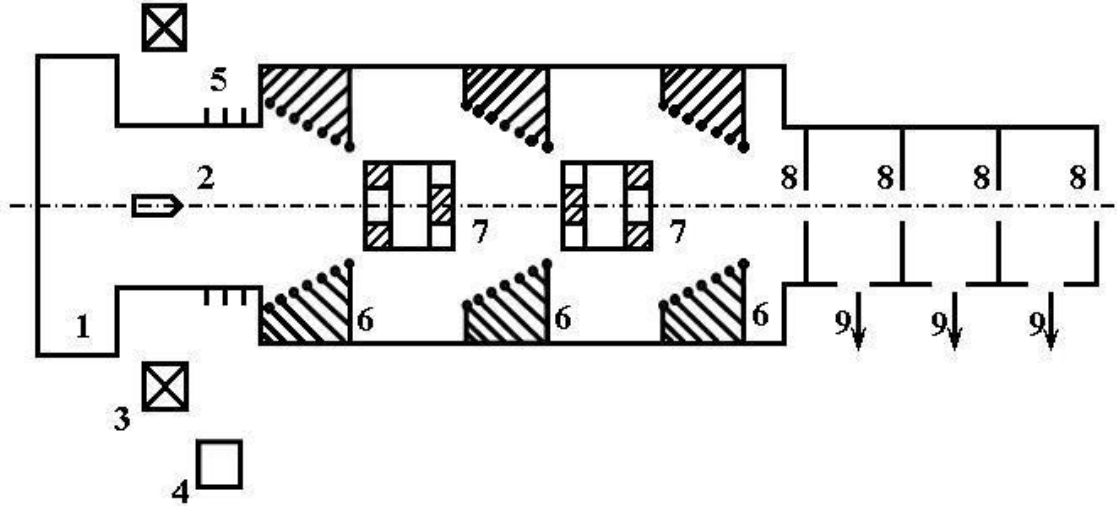

Fig. 1. (1) feeding cassette, (2) particles, (3) pulsed magnetic field coil,
(4) electron gun, (5) electrostatic particle acceleration area,
(6) - spiral waveguide sections, (7) doublets of quadrupole electrostatic lenses,
(8) pulse diaphragms, (9) individual evacuation of buffer cavities.

**Application**

*1. Ballistics*

Let us calculate the motion of the particle accelerated by the electrostatic fieldwith air resistance taken into account. The equation of motion of the particle can be written without allowance for the gravitational attraction to the Earth

$$mdV/dt = -\rho C_x S_{tr} V^2 / 2, \qquad (20)$$

where m is the mass of the particle, V is the velocity, $\rho = \rho_0 e^{-z/H0}$ is the barometric formula for the atmospheric density change with altitude, $\rho_0 = 1.3 * 10^{-3}$ g/cm$^3$ is the air density at the surface of the Earth, $H_0 = 7$ km is the altitude at which the density drops by a factor e. In our case, the particle is a cylinder, $S_{tr}$ is the cross-section of the particle, $S_{tr} = \pi d^2 / 4 = 2.8 * 10^{-7}$ cm$^2$, and $C_x$ is the aerodynamic coefficient.

The aerodynamic coefficient, or the drag coefficient, is a dimensionless quantity that takes into account the "quality" of the particle form

$$C_x = F_x / (½) \rho_0 V_0^2 S_{tr}. \qquad (21)$$

Equation (20) can be written as

$$V(t) = V_0 / [1 + \rho C_x V_0 * S_{tr} * t/2m]. \qquad (22)$$

To calculate the particle velocity variation in time, it is necessary to find the aerodynamic coefficient $C_x$.

*2. Calculation of the particle drag coefficient in the air*

We assume that the particle has the shape of a cylindrical rod with a conical head. When a nitrogen molecule hits the cone, the change in the longitudinal velocity of the molecules is

$$\Delta V_x = V_x * \Theta_h^2 / 2, \qquad (23)$$

where $\Theta_h$ is the cone angle. The momentum transferred by the gas molecules to the macroparticle is

$$p = mV = \rho V_x S_{tr} t * \Delta V_x. \qquad (24)$$

The change in the momentum per unit time is a force, namely, a drag force.

$$F_{x1} = (½) \rho V_x S_{tr} * V_x * \Theta_h^2. \qquad (25)$$

Dividing Fx1 by $(½) \rho V_x^2 S_{tr}$, we obtain the drag coefficient for a pointed cone at the mirror reflection of molecules from the cone (the Newton formula)

$$C_{x\,air} = \Theta_h^2. \qquad (26)$$

Let the length of the conical part of the particles be $l_{cone} = 3$ mm. This means that the cone angle is $\Theta_h = 2 * 10^{-3}$ and $C_{x\,air} = 4 * 10^{-6}$.

*3. Passage of particles through the atmosphere*

Let us calculate the passage of the particle with a drag coefficient $C_{x\ air} = 4 * 10^{-6}$ through the Earth's atmosphere. The decrease inthe velocity with time due to air resistance is described by (20). We choose the following parameters: $\rho = \rho_0 e^{-z/H0}$, where $H_0 = 7$ km, $C_{x\ air} = 4 * 10^{-6}$, $S_{tr} = 3 * 10^{-7}$ cm$^2$, $V_{fin} = 10^6$ cm / s, m = $6 * 10^{-6}$ g, and $\rho_0 = 1.3 * 10^{-3}$ g/cm$^3$. For the first few seconds of flight the decrease in the velocity is very small.

Further, as the altitude increases, the contribution of this term will decrease exponentially. Thus, for the altitude z = 2 km the factor $e^{z/H0}$ is 0.75, for z = 4 km it is 0.57 etc. This means that the velocity loss of the particle during its ascent and descent can be neglected and it can be considered as flying in vacuum.

*4. Particle penetration depth in aluminum*

The depth to which a tungsten wire segment moving at a velocity of $V_0 = 10$ km / s penetrates in aluminum can be estimated in the following way. Since the processes in question are fast, it is necessary to consider only interaction with the atoms that are on the way of the wire segment.

Let us find the energy to be spent for heating one centimeter of aluminum with the cross sectional area equal to the cross-section of the tungsten wire $S_{tr} = 2.8 * 10^{-7}$ cm$^2$ from room temperature to the melting point. Then we should add the solid–liquid phase transition energy to this energy. The fluid should be heated to the evaporation temperature and the liquid–vapor phase transition energy should be taken into account.

The total expenditure of energy is $\Delta W_{1\ cm} = 10^{-2}$ J / cm.

The kinetic energy of the tungsten wire segment with a diameter of $d_{sh} = 6$ µ, length $l_{sh} = 1$ cm, and velocity $V_0 = 10$ km / sis $W_{sh} = 0.3$ J. the analysis of collisions of meteorites with solids allows the conclusion that at these velocities about 40% of the kinetic energy of meteorites is spent for heating and evaporation of the target material.
We assume that out of the entire kinetic energy of the tungsten wire segment $W_{sh} = 0.3$ J , the energy spent for evaporation of aluminum is $W_{sh1} = 0.1$ J.

Now, dividing $W_{sh1}$ by the specific energy loss $\Delta W_{1\,cm}$, we find the depth of penetration of the segment in aluminum: $L_{Al} = W_{sh1} / \Delta W_{1\,cm} = 10$ cm.

*5. Drag-affected penetration depth of particles in water*

Let us find the drag force acting on the particle moving in water. Unlike the case in the air, reflection of molecules from the particle in water should be considered as being diffuse, which leads to the drag coefficient $C_{x\,w} \approx 1$. The time dependence of the veclocity reduction will be described by (23). We denote the particle stopping velocity in water by $V_s$. Let $V_s$ be 0.1 km / s. From (23) we obtain the particle slowing-down length

$$l_s = \int_0^{t_0} V(t)\,dt = [2m / (\rho_w * C_{x\,w} * S_{tr})] * \ln(1 + V_0/V_s), \qquad (27)$$

where $V_0$ is the velocity at which the particle enters water, $\rho_w = 1$ g/cm$^3$ is the density of water, $C_{x\,w} = 1$, m is the mass of the particle, and $V_s = 0.1$ km / s. Thus, the particle slowing-down length in water is

$$l_{s1} = \{2m / (\rho_w * C_{x\,w} * S_{tr})\} * \ln(1 + V_0/V_s), \qquad (28)$$

The coefficient of the factor $\ln(V_0/V_s - 1)$ has the dimension of length and is 40 cm for the parameters chosen above. In our case, $V_0/V_s = 10^2$, and relation (28) yields the drag-affected particle penetration depth in water $l_{s1} = 180$ cm. Contributions to particle slowing-down in water will also come from the other processes, e.g., viscosity of water.

*6. Penetration depth of particles in water as affected by water viscosity*

The equation of motion of the particle entering water can be written as

$$m dV_x / dt = -F_{x2} \qquad (29)$$

The drag force due to viscosity is

$$F_{x2} = \sigma S = \eta (dV/dr) * 2\pi r_s h\, l_{sh} = \eta (V/\delta) * 2\pi r_{sl}, \qquad (30)$$

where $\eta = 10^{-2}$ Poise is water viscosity, $r_{sh} = 3 * 10^{-4}$ cm is the radius of the particle, $l_{sh} = 1$ cm is the particle length, $dV/dr$ is the radial gradient of the longitudinal water velocity near the particle, and $\delta$ is the characteristic length of the velocity variation in radius.

Equation (29) with the viscosity drag force in the form of (30) can be rearranged in the form

$$dV / V = - (\eta / m\delta) * 2\pi r_{sh} l_{sh} * dt, \qquad (31)$$

which has the solution

$$V = V_0 \exp [- (\eta / m\delta) * 2\pi r_{sh} l_{sh} * t]. \qquad (32)$$

Now we can find the path traveled by the macroparticle to a stop in water

$$l_{water} = \int_0^\infty V dt = V_0 (m\delta / \eta * 2\pi r_{sh} l_{sh}). \qquad (33)$$

To find the particle penetration depth in water, we need to find $\delta$, the characteristic length of the ramp near the water velocity variation in radius near the particle.

Let us find it from the Navier equation

$$\partial V_x / \partial t + V_x \partial V_x / \partial x = (\eta / \rho r) d / dr (r dV_x / dr) + \partial^2 V_x / \partial x^2. \qquad (34)$$

First we will see in which case the term $V_x \partial V_x / \partial x$ is much greater than $(\eta / \rho) \partial^2 V_x / \partial x^2$. Let $\partial x = l_{ch}$ be the characteristic length over which the velocity varies. Then $V_x \partial V_x / \partial x = V_x^2 / l_{ch}$, $(\eta / \rho) \partial^2 V_x / \partial x^2 = (\eta / \rho) V_x / l_{ch}^2$. The condition $V_x \partial V_x / \partial x \gg (\eta / \rho) \partial^2 V_x / \partial x^2$ is satisfied if $\rho V_x l_{ch} / \eta = Re \gg 1$, where Re is the Reynolds number and the relation $Re \gg 1$ is certainly satisfied in our case.

Now we rearrange (34) in the form

$$\partial V_x / \partial t + V_x \partial V_x / \partial x = dV_x / dt = (\eta / \rho r) d / dr (r dV_x / dr), \qquad (35)$$

where we replaced $\partial / \partial t + V_x \partial / \partial x$, the partial derivative with respect to time and coordinate by $d / dt$, the total time derivative.

Substituting $dV_x / dt = -(\eta / m) * 2\pi r_{sh} l_{sh} * (dV_x / dr)$ found from (33) into the Navier equation, we obtain the equation

$$-(\eta/m)*2\pi r_{sh} l_{sh} *(dV_x/dr) = (\eta/\rho r) d/dr(rdV_x/dr), \qquad (36)$$

which, after cancellation of $\eta$ (the water viscosity) and transpositions, takes the form

$$2\pi r_{sh} l_{sh} \rho / m * (rdV_x / dr) = -d/dr (rdV_x / dr). \qquad (37)$$

Replacing $rdV_x / dr$ by y, we get:

$$d (\ln y) = -(2\pi r_{sh} \text{ sh } \rho / m) dt, \qquad (38)$$

or

$$dV_x / dr = (C_{1v} / r) \exp(-2\pi r_{sh} l_{sh} \rho r / m). \qquad (39)$$

To find the integration constant $C_{1v}$, we integrate (39) again and obtain

$$V = C_{1v} * \text{Ei} (2\pi r_{sh} l_{sh} \rho r / m) + C_{2v}, \qquad (40)$$

where $\text{Ei}(z) = \int_0^\infty \exp(-t) dt / t$ - expint (z) is the exponential integral. Here we made a replacement $2\pi r_{sh} l_{sh} \rho r / m = t$; the limits of integration with respect to r are from $r = r_{sh}$ to infinity, and the lower limit of integration with respect to t is accordingly $t = 2\pi r^2_{sh} l_{sh} \rho / m = 2\rho_{water}/\rho_{tungsten} = 0.1$.

The function $\text{Ei}(z)$ cannot be expressed in terms of elementary functions; it can be approximated as [5]

$$(\tfrac{1}{2}) e^{-z} \ln(1 + 2/z) < \text{Ei}(z) < e^{-z} \ln(1 + 1/z), \quad z > 0, \qquad (41)$$

Equation (40) describes the radial dependence of the velocity of the medium as the particle moves in it at a velocity V. At large distances from the particle $V = 0$, and thus $C_{2v} = 0$. The boundary condition on the surface of a body is usually taken to be the condition of "sticking" of the medium particles to the body, i.e, when $r = r_{sh}$, the velocity of the medium is $V = V_0$, from which we can find the constant $C_{1v}$

$$V_0 = C_{1v} * Ei(0.1) \text{ and}$$

$$C_{1v} = V_0/Ei(0.1). \tag{42}$$

Expression (32) for the velocity gradient at the surface of particles can now be written as

$$dV_x / dr = V_0 / [r_{sh} Ei(0.1) * \exp(0.1)]. \tag{43}$$

Thus, the characteristic amount of the radial velocity variation δ is

$$\delta = r_{sh} * Ei(0.1) * \exp(0.1), \tag{44}$$

and, according to formula (33), the particle path length particles in water is

$$l_{water} = \int_0^\infty Vdt = V_0(m\delta/\eta * 2\pi r_{sh} l_{sh}) = V_0[m * Ei(0.1) * \exp(0.1) / 2\pi\eta l]. \tag{45}$$

Finally,

$$l_{water} = (V_0 m/2\pi\eta l_{sh}) * Ei(0.1) * \exp(0.1). \tag{46}$$

Substituting $V_0 = 10$ km/s, $m = 6 * 10^{-6}$ g, $\eta = 10^{-2}$ Poise, $l_{sh} = 1$ cm, and $Ei(0.1) = 2.6$, $\exp(0.1) = 1.1$ into (46), we obtain

$$l_{water} = 230 \text{ cm}. \tag{47}$$

To find the penetration depth in water as affected by two processes, viscous drag and friction, we should be add the inverse values and find the penetration depth $L_{water}$ by the formula

$$L_{water} = l_{s1} * l_{water} / (l_{s1} + l_{water}) = 1m. \tag{48}$$

Thus, the penetration depth of the tungsten wire segment in water is approximately $L_{water} = 1m$.

**Conclusion**

Even a weakly magnetized particle will rotate in the magnetic field of the Earth like an ordinary compass needle. If the particle turns relative to the velocity vector, all aerodynamics and penetration depth formulas will be incorrect, which means that we must look for other ways to orient macroparticles relative to the axis of acceleration.

If we apply four sectors of a material with different resonant absorption frequencies to the macroparticle, the particle motion can be controlled by evaporation of the material from the corresponding sector and transverse ablative acceleration of the particle. Considering that the velocity of the evaporated material is $V_{ab} = 0.1$ km / s and the mass of the evaporated material is $m_{ab} = 0.1$ m (0.1 of the wire segment mass), we find that in this way we can get angles of deviation from the original direction on the order of $\theta_\perp = 10^{-3}$.